\documentclass{PoS}

\title{Galactic chemical evolution: the observational side}

\ShortTitle{Galactic chemical evolution: the observational side}

\author{\speaker{Andrew McWilliam}\\
       Carnegie Observatories\\
       813 Santa Barbara Street\\
       Pasadena, CA 91101, USA\\
       E-mail: \email{andy@obs.carnegiescience.edu}}

\abstract{In this review I first outline some interesting ideas in chemical
evolution, necessary for understanding the evolution of galaxies from measured elemental
abundance ratios.  I then discuss abundance results from studies of stars in Local Group dwarf
galaxies and the globular cluster Omega Cen.  Finally, I present a qualitative
scenario of prolonged chemical evolution in a leaky box that can explain the observed
abundance ratios trends in these dwarf systems.

Unfortunately, space limitations prevent a comprehensive discussion of the vast field of
observational chemical evolution, so I have limited this paper to a few interesting subtopics.
I have completely omitted the Galactic bulge; this may be as well, because there is
some disagreement over the measured abundance ratios, except for [O/Fe] which follows the thick
disk trend.}

\FullConference{11th Symposium on Nuclei in the Cosmos\\
                19-23 July 2010 \\
                Heidelberg, Germany.}

\begin{document}

%\section{Introduction}

%reasons for studying chemical evolution
%
%-->CE and yields solved simultaneously  
   %--> probe the full range of environments w/RGB stars
   %--> timescale, SFR, IMF, inflows/outflow, accretion
%
%***Yes, the observational side, but in comparison with simple predictions

%I was asked to review chemical evolution.  Unfortunately, time limitations in
%the talk and page limitations here prevent a comprehensive discussion, so I
%present a limited presentation of some interesting subtopics.  I have completely
%omitted the discussion of the Galactic bulge; this may be as well, since there
%is some disagreement over the measured abundance ratios, except now [O/Fe].

\section{An Introduction to Chemical Evolution}

This paper is mostly concerned with the observational side of chemical evolution;
however, in order to interpret the measured abundance ratios it is necessary
to understand some basic qualitative concepts in chemical evolution theory.

In the 1950s both observations and theory (Chamberlain \& Aller 1951; 
Hoyle 1955; Burbidge et al. 1957; Preston 1959) indicated that most of the
chemical elements were produced by stars and that the Galaxy had undergone
chemical evolution due to nuclear processing by many generations of stars.

\subsection{The Simple Model}

Perhaps the most basic chemical evolution model is the Simple Model of van den Bergh (1962)
and Schmidt (1963.)  This idealized scenario begins with a mass
of metal-free gas, and a constant rate of star formation (SFR), in stellar generations,
with a fixed distribution of stellar masses, or initial mass function (IMF).  Each stellar 
generation produces an identical mass of metals from its massive stars, and fossil remnants
from the dwarfs; the metals are instantaneously, and homogeneously, mixed
into the unused gas, thus increasing the gas metal abundance.  
There is no inflow or outflow of material to the region, i.e., it is a ``closed box''.  
Searle \& Sargent (1972) introduced the concept of {\em ``yield''}, which is the ratio of the mass of
metals produced to the mass locked-up in low mass stars.  In a Simple Model that runs to gas
exhaustion, the average metallicity of the dwarfs is equal to the {\em ``yield''}.  
Thus, the average of the metallicity distribution function (MDF) for a stellar
system indicates the yield of metals from its stars.

The Simple Model 
age-metallicity relation is linear, while the MDF is logarithmic: for example, there are a
thousand stars at [Fe/H]=0 for each star at [Fe/H]=$-$3.
The observed solar neighborhood MDF shows a deficiency in metal-poor stars compared to
the Simple Model; this deficit is known as the {\em ``G-dwarf problem''}.  
The widely accepted explanation for the G-dwarf problem
is that inflow of low-metallicity gas has occurred during the evolution of the
Galactic disk, with an exponential decay timescale of $\sim$5 Gyr (e.g., Sommer-Larsen 1991).
At early times there was relatively little
gas present so only small numbers of metal-poor G-dwarf stars
were produced, compared to later epochs when the metallicity was higher.  

%Enough stars, and metals, were produced during the
%inflow such that the inflowing gas did not reduce the total gas metallicity.  

Chemical evolution models with increasing complexity have
been constructed, and they are now much more realistic; for discussion of these
I refer the reader to the talk by Nick Prantzos at this conference.  However, I find the Simple Model
a useful heuristic tool and a good starting point for thinking about ideas in chemical evolution.

\subsection{The Metallicity Distribution Function}

While the mean metallicity of a Simple Model can be used to determine the yield, in
stellar systems the mean of the MDF can be affected in several ways:

The mean [Fe/H] for the Galactic halo, found by Hartwick (1976), is near $-$1.6 dex,
which he explained as due to the halo losing its gas
before chemical evolution could go to completion; thus, {\em gas outflow} terminated
halo evolution at low metallicity.

This contrasts with the explanation for the metallicities of gas and young stars in the LMC
and SMC, at [Fe/H]=$-$0.3 and $-$0.6 dex, respectively (e.g., Russell \& Dopita 1990).  The 
presence of young and old stars (supergiants, RGB and carbon stars) and high gas fractions
suggest prolonged evolution, and a low SFR, for these dwarf galaxies; thus, the
metallicity may not have had time to reach solar values.  However, outflows could
also explain the low [Fe/H] of the LMC and SMC;
realistic models are required for a full understanding.

Changes in the IMF can also affect the mean metallicity: the yield depends on the
mass of iron produced by massive stars, and the mass locked-up in low-mass dwarf stars,
both of which varies with the IMF.  In this way Ballero et al. (2007) employed an IMF
weighted to massive stars to explain how rapid bulge formation, with no Fe from SNIa,
could produce a high mean metallicity, near the solar value.  An alternative, suggested
by A.Pipino, is that the bulge experienced huge mass inflows early-on, giving an extreme
G-dwarf problem and higher mean [Fe/H].

\section{Alpha Elements}

Wallerstein (1962) and Conti (1967) first recognized the factor of two 
enhancements of Mg, Si, Ca, Ti and O in Galactic halo RGB stars.  These even-numbered
elements, from O to Ti, are referred to as $\alpha$-elements, although no
single nuclear reaction is responsible for their synthesis.  Plots of [$\alpha$/Fe]
versus [Fe/H] in the solar neighborhood show a plateau below [Fe/H]$\sim$$-$1, followed
by a steady decline to [$\alpha$/Fe]$\sim$0 at [Fe/H]$\sim$0.  The trend is, actually, a composite
of the ratios transitioning from halo, to thick disk and thin disk populations.

%Studies of the composition
%of Galactic halo red giant stars by Wallerstein (1962, 1963) and Conti (1967) found that
%even-numbered light elements (e.g. Ca, Ti, O) are over-abundant in the halo,
%relative to the solar composition.  These elements have become known as
%{\em ``alpha elements''}, although there is no single nuclear reaction chain 
%responsible for their synthesis.  Amongst the light elements O, Mg, Si, S, Ca, and Ti
%are now known to follow a similar trend in the solar neighborhood, a roughly factor of
%two enhancement in [$\alpha$/Fe] below [Fe/H]$\sim$$-$1, with a decline toward
%solar composition at higher [Fe/H].  It is important to note that these trends
%depend on the stellar population under consideration (e.g., thin disk, thick disk, halo).

Tinsley (1979) proposed that the trend of [O/Fe] with [Fe/H] resulted from the time
delay between SNII and SNIa.
At early times, and low [Fe/H], the SNII [O/Fe] values prevailed; subsequent addition of
SNIa lowered the [O/Fe] ratio because SNIa produce Fe but no O.  A Simple Model linear age-metallicity 
relation and the observed [O/Fe] decline, beginning near [Fe/H]$\sim$$-$1, indicates a
$\sim$1 Gyr delay for the onset of significant SNIa Fe production, roughly consistent
with detailed calculations of Matteucci \& Greggio (1986).
I note that while the most massive SNIa progenitors have delays $\sim$0.1 Gyr, a significant effect
on [O/Fe] does not occur until much later.
%The most rapid SNIa occur in $\sim$0.1 Gyr, but the net  
%are a minority of the total population and must
%do not make enough Fe to significantly affect the Galactic [O/Fe] ratio.

%Tinsley (1979) proposed an explanation for the trend of [O/Fe] with [Fe/H] in the
%solar neighborhood, involving the time delay between core-collapse, Type~II
%supernovae (SNII) and the binary mass-transfer Type~Ia supernovae (SNIa).  At
%early times, and low [Fe/H], the massive envelopes of SNII progenitors were
%responsible for enhanced [O/Fe], while at later times, and higher [Fe/H], SNIa
%produced additional Fe, but not O, thereby reducing the [O/Fe] ratio with increasing
%[Fe/H].

The calculations of Matteucci \& Brocato (1990, henceforth MB90), adapted in Figure~1, showed
that the down-turn,
or knee, in the [O/Fe] versus [Fe/H] plot should depend on the SFR.  High SFR systems, like 
bulges and giant ellipticals, reach high [Fe/H] before the SNIa time-delay, and the decline
in [O/Fe]; similarly,
low SFR systems, such as dwarf galaxies, should show a decline in [O/Fe] at low [Fe/H].

MB90's strict age-metallicity relation (probably unrealistic), meant that only the most 
massive stars contributed to the composition at the lowest metallicity.
%
%MB90's strict age-metallicity relation (probably unrealistic), meant that at
%metal-poor times the mean SNII progenitor mass was higher than at later times.  
%
This is the cause for MB90's gentle slope to higher [O/Fe] below the knee, because the
O/Fe yield ratio increases with increasing SNII mass.  Massive SNII produce higher [O/Fe]
due to their larger envelopes, where O is produced, and because of greater fallback of Fe, than
lower mass SNII.  Following this idea Wyse \& Gilmore (1993) showed that shallower 
IMF slopes (i.e., weighted to massive stars) produce larger [O/Fe] ratios.  Thus, the 
[O/Fe] ratios below the knee can constrain the IMF slope, while the [Fe/H] of the knee 
can constrain the SFR (and formation timescale) for a stellar system (see Figure~1).

The alpha elements can be divided into two categories: {\em hydrostatic} (O and Mg), made
in the envelopes of massive SNII progenitors before the explosive event, and 
{\em explosive} (Si, S, Ca, and Ti), principally made during the explosion.  Supernova nucleosynthesis
calculations (e.g., Woosley \& Weaver 1995) show that O and Mg are made 
by progenitors with masses $\sim$35M$_{\odot}$, while the production of Si through Ca peaks near
20--25M$_{\odot}$ SNII progenitors.  Thus, the ratio of [Mg/Ca] should be sensitive to the IMF slope.

The O yield from massive stars is also predicted to be sensitive to metallicity (e.g., Maeder 1992),
due to the stripping of the envelope by stellar winds, related to the Wolf-Rayet phenomena.  In this
way the [C/O] yield ratio increases with [Fe/H], since carbon that would normally be burned to oxygen
is removed from the star. This mechanism appears to have operated in the Galactic thick disk and bulge
(Cesctti et al. 2009).  Wolf-Rayet stars produced via envelope stripping from binary mass transfer 
should also decrease the oxygen yield and increase the [C/O] ratio.

Thin disk and Galactic bulge [O/Mg] trends show a steep decline above [Fe/H]$\sim$$-$1 (see McWilliam
et al. 2008).  While the bulge results are now not agreed upon by all observers, there is no disagreement
for the thin disk trend.  The [O/Mg] decline in the disk can result from a decrease in oxygen yields
from massive stars due to envelope stripping by metallicity-dependent stellar winds
(e.g., Cescutti et al. 2009).

While predicted SNIa yields make effectively no O or Mg, they do make significant amounts of
Ca (e.g., Nomoto 1984), near the solar [Ca/Fe] ratio.  Thus, Tinsley's scenario suggests
a smaller amplitude change in [Ca/Fe] than [O/Fe].

%\begin{figure*}[ht]
\begin{figure}[ht]
\centering
\includegraphics[angle=0,width=12cm]{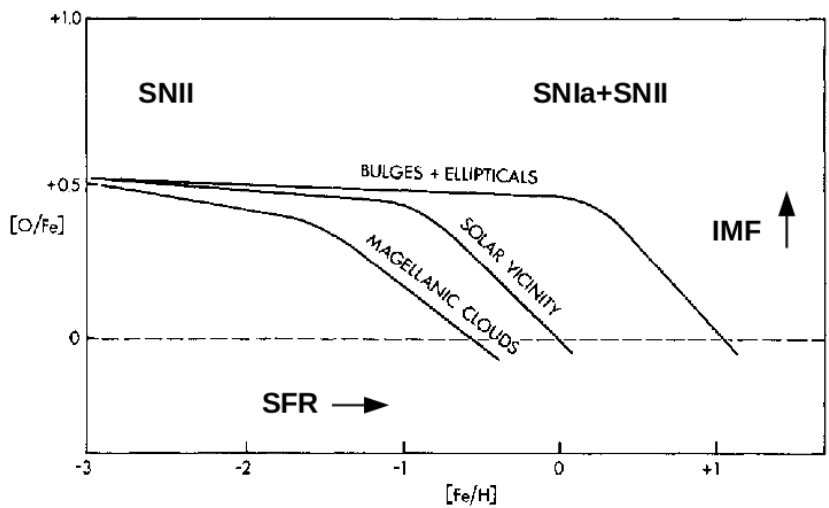}
\caption{Predicted [O/Fe] versus [Fe/H] for systems with different
SFR, modified from Matteucci \& Brocato (1990).}
%\label{cmd_b6}
%\end{figure*}
\end{figure}

%An important point to note in Figure~1 is
%that the nearly constant [O/Fe] ratio at lower [Fe/H] than the knee, is not constant,
%but slopes slightly downward.  The reason is that at very early times only the highest
%mass SNII progenitors would have evolved, and higher mass SNII produce more O in their
%massive envelopes than low mass SNII.  This dependence of the [O/Fe] ratio is expanded
%upon by Wyse \& Gilmore (1992, 1993) who show that shallower IMF slopes (i.e., weighted
%toward more massive stars) produce larger [O/Fe] ratios.  Thus, it appears that the
%trend of [O/Fe] with [Fe/H] can, in principle, be used to estimate the history of the
%SFR and IMF slopes of massive stars in a stellar system.

\section{The Chemical Composition of Dwarf Galaxies}

\subsection{The Magellanic Clouds}

%The earliest abundance studies of LMC stars were for supergiants (e.g., Spite?? 198?;
%Russell \& Bessell 1989?; McWilliam \& Williams 1989; Hill???) as these are bright
%enough, at V$\sim$13, for 4m class telescopes.  However, supergiant stars are so massive 
%that they probe relatively young material, and not the ancient history of chemical
%evolution; also, problems with non-LTE effects cast doubt on the reliability of the
%supergiant results (e.g., McWilliam \& Williams 1990?).

Early abundance studies of Magellanic Cloud stars were restricted to bright supergiants (e.g.,
Spite et al. 1986; Russell \& Bessell 1989; McWilliam \& Williams 1989),
which only traced the very recent composition; typically, the supergiants were hot, 
with subsequent uncertainty about non-LTE effects on the results.

Later abundance studies of LMC red giant branch (RGB) stars
probed all ages; however, initial results from Hill et al. (1995) and Korn et al. (2002)
were too few to discern a trend of [O/Fe] with [Fe/H].  Smith et al. (2002),
with 12 LMC RGB stars, plus the earlier results, showed a steep decline in
[O/Fe] well below that for the Milky Way, with [O/Fe] lower by almost 0.3 dex
at the metal-rich end (near [Fe/H]=$-$0.2 dex).  It was not clear whether the
[O/Fe] ratios at the low [Fe/H] end are the same in the LMC and Galaxy; this zero-point
issue should be resolved.

%Initially, too few RGB stars were studied to make a
%reliable comparison with the [O/Fe] versus [Fe/H] trend in the Milky Way.  The first
%persuasive result came from Smith et al. (2002), who had a dozen LMC RGB stars, 
%supplemented with values from Hill et al. (1995) and Korn et al. (2002).

%Later studies, with larger telescopes, reached LMC red giant branch (RGB) stars, that
%cover the full range of ages.  Initially, too few RGB stars were studied to make a
%reliable comparison with the [O/Fe] versus [Fe/H] trend in the Milky Way.  The first
%persuasive result came from Smith et al. (2002), who had a dozen LMC RGB stars, 
%supplemented with values from Hill et al. (1995) and Korn et al. (2002).
%
%\begin{figure}
%***Show the Smith et al. (2002) Figure for [O/Fe] vs. [Fe/H] here
%\end{figure}
%

%In Figure~2 I show the Smith et al. (2002) results, plus literature values, for [O/Fe]
%in the LMC, compared to the Milky Way.  The LMC [O/Fe] seems significantly lower than
%the Milky Way, almost 0.3 dex, at the metal-rich end, near [Fe/H]=$-$0.2 dex.  One 
%concern is the paucity of LMC points at [Fe/H]=$-$1 and below; in particular, it is
%not clear whether the low metallicity ends are the same in both systems.  This is 
%important to improve upon in order to eliminate systematic effects as a cause for
%the difference with the Milky Way [O/Fe] trend.  Still, the Smith et al. (2002) result
%shows a very steep, and steepening [O/Fe] trend at higher [Fe/H], in contrast to an
%apparent flattening-out in the Milky Way.

A large sample of LMC RGB stars studied by Pomp\'eia et al. (2008) provided [$\alpha$/Fe]
trends for O, Mg, Ca, Si and Ti.  The handful of O/Fe abundance ratios 
overlapped with Smith et al. (2002).  Relative to the Galaxy the LMC [Ca/Fe] and [Ti/Fe] ratios
of Pomp\'eia et al. (2008) are deficient at all [Fe/H]; remarkably, [Si/Fe]$\sim$0 for all [Fe/H].
The average [Mg/Fe] LMC trend is deficient, relative to the Milky Way, below [Fe/H]=$-$0.6, but 
could be described as bimodal, as if halo and thin disk populations overlapped in [Fe/H].
%
%but the [Mg/Fe] trend almost appears bimodal, as if the low thin disk value
%exist above [Fe/H]=$-$1 simultaneously with an enhanced, thick disk, ratio below
%[Fe/H]=$-$0.6; certainly, above [Fe/H]=$-$0.5 dex the LMC and Milky Way [Mg/Fe] ratios 
%appear identical.
%
Thus, there is rough confirmation that [O/Fe] and the other [$\alpha$/Fe] ratios are
deficient in the LMC, relative to the solar neighborhood, as predicted by MB90, but
further investigation is warranted.  

%However, there is some evidence that alpha-elements do not all behave the same
%way with [Fe/H].

\subsection{Alpha Elements in Local Group Dwarf Galaxies}

The first Local Group dwarf galaxy abundance measurements, for Draco, Ursa Minor and
Sextans, by Shetrone et al. (2001), gave an average of Mg/Fe, Ca,/Fe and Ti,/Fe
$\sim$0.2 dex deficient, relative to the Milky Way (O was not measured),
qualitatively consistent with the predictions of MB90.

%High resolution abundance measurements of stars in the
%Draco, Ursa Minor and Sextans Local Group
%dwarf galaxies was first attempted by Shetrone et al. (2001);
%only a handful of stars were studied, so the results for all
%three galaxies were combined.  The abundances of Ca, Ti and Mg were
%measured and averaged, for an estimate of [$\alpha$/Fe].  A comparison
%with the Milky Way showed that these three dwarf galaxies were
%deficient in alpha elements by$\sim$0.2 dex, as expected from the
%MB90 predictions.  Notably, the few oxygen points did not provide a
%clear-cut difference between [O/Fe] in these galaxies and the Milky Way.

Shetrone's results provoked an important question: if the halo is composed of
accreted dwarf galaxy fragments, then why does it possess higher
[$\alpha$/Fe] than the Local Group dwarf galaxies?  The answer is that the
halo must be made mostly of early dwarf galaxy fragments, which had not yet
suffered significant enrichment by SNIa when they were accreted.
Not withstanding, there are plenty of examples of Galactic halo 
stars with low [$\alpha$/Fe] (e.g. Nissen \& Schuster 1997; Brown et al. 1997).

For the more metal-rich stars in the Sagittarius dwarf spheroidal galaxy 
(henceforth Sgr dSph) [$\alpha$/Fe] deficiencies were found
by Smecker-Hane \& McWilliam (2002), McWilliam \& Smecker-Hane (2005a), 
Sbordone et al. (2007), and Carretta et al. (2010).  However, the metal-poor 
Sgr dSph stars possessed normal halo $\alpha$-element enhancements.  Similar 
results for the [O/Fe] ratios in the Sculptor dSph were found by Geisler et al.
(2005); and Letarte et al. (2010) recently found low [$\alpha$/Fe], for Mg, Si,
Ca and Ti, in the Fornax dSph, although no O measurement was made.
Cohen \& Huang (2009,2010) found low metallicity knees in the [$\alpha$/Fe] trends
for the Draco and Ursa Minor dSphs, and claimed halo-like ratios for the most metal-poor
members of these dwarf galaxies. 

These abundance results are qualitatively consistent with Tinsley's time-delay scenario
and the predictions of MB90.
They are also consistent with the idea that the halo was formed from
accreted dwarfs at very early times, before the bulk of SNIa had occurred.
Subsequent enrichment of present day dwarf galaxies decreased the [$\alpha$/Fe]
ratios, but the older, metal-poor, stars in the nearby dSphs have halo-like
compositions.

\vfill\eject

\subsection{Sodium and Aluminium}

%Aluminium and Sodium deficiencies of $\sim$0.3 dex were found for the
%Sgr dSph by Smecker-Hane \& McWilliam (2002?), in agreement with a
%result for two stars found by Bonifacio et al. (2000).  More recently,
%Sbordone et al. (2007) has confirmed the Al and Na deficiencies, but
%the results of Carretta et al. (2010) are ambiguous for these two elements.

Bonifacio et al. (2000), Smecker-Hane \& McWilliam (2002), and Sbordone et al. (2007)
found [Al/Fe] and [Na/Fe] deficiencies of $\sim$0.3--0.4 dex in the Sgr dSph;
results of Carretta et al. (2010) are inconclusive.

For the LMC RGB stars Smith et al. (2002) found a mean [Na/Fe]=$-$0.3 dex, similar
to the results from F--G supergiants by Hill et al. (1995).
Pomp\'eia et al. (2008) found Na/Fe deficiencies of$\sim$0.3--0.4 dex for her large sample of
LMC RGB stars, relative to the solar neighborhood.  For the Draco and UMi dSphs Cohen \& Huang 
(2009, 2010) found Na/Fe deficiencies, relative to the Galaxy, for the metal-rich end of their 
sample, similar to their $\alpha$-element deficiencies.  Large [Na/Fe] deficiencies, $\sim$$-$0.7 dex
were found for the stars in the Fornax dSph, by Letarte et al. (2010), for which the mean
[Fe/H] was $-$0.8 dex.  For the Sculptor dSph Geisler et al. (2005) found mean [Na/Fe]
and [Al/Fe] near $-$0.5 and $-$0.4 dex, respectively, similar to the deficiencies found
in the Sgr dSph.  Thus, it appears that the Na deficiencies, at least, are common-place
among the dwarf galaxies, and that when measured Al is also deficient.  More work on Al/Fe
ratios would be useful.

The main source for Al and Na is thought to be the hydrostatic phase of SNII progenitors
(e.g., Woosley \& Weaver 1995), modulated by the neutron excess (i.e., metallicity)
as described by Arnett (1971).  Thus, Al and Na should behave much like
alpha-elements, but with a metallicity-dependent yield trend imposed; so, it is not surprising
that Al and Na are deficient in dwarf galaxies, where alpha-element deficiencies are also observed.

%The main production of Al and Na is thought to occur in the hydrostatic phase of
%massive stars that end as Type~II SNe (e.g. WW95?), although 
%modulated by the neutron excess (e.g. Arnett 1971), being reduced at low
%metallicities.  Other sources of Al and Na include hot bottom envelope 
%proton burning in evolved stars; while such self-pollution my dominate the Al
%abundance in the envelope of an individual evolved star it is likely a minor
%contributor to the Galactic Al content.  Thus, Al and Na are mostly made by
%massive stars, similar to O, and it might be expected that the behavior of [Al/Fe]
%and [Na/Fe] should be similar to [O/Fe], but also increasing with increasing [Fe/H].
%In this way one could group Al and Na with the alpha elements.  Indeed, it is not
%then surprising that the $\alpha$-element deficiencies in dwarf galaxies seem
%to be accompanied by Al and Na deficiencies.

\begin{figure}[ht]
\centering
\includegraphics[angle=0,width=9cm]{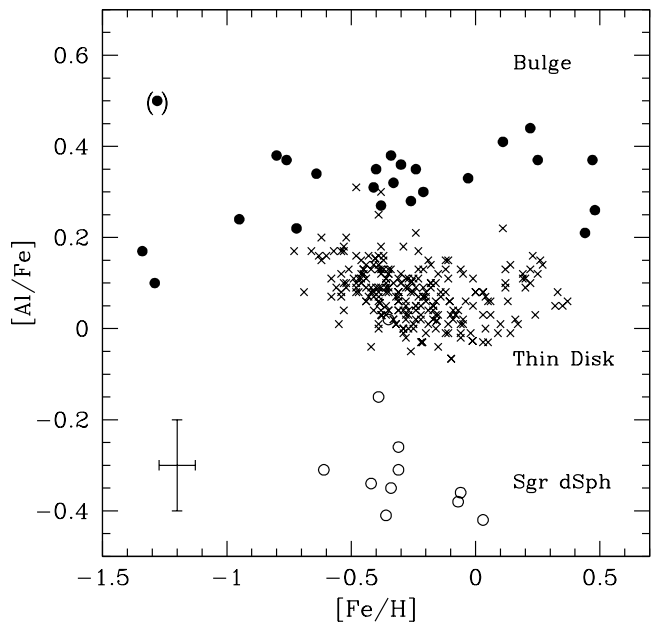}
\caption{[Al/Fe] for the Sgr dSph, Galactic thin disk, and the Galactic bulge, adapted from
Fulbright et al. (2007).}
\end{figure}

In Figure~2 I show [Al/Fe] versus [Fe/H] for three systems: Sgr dSph, the thin
disk, and the Galactic bulge, from Fulbright et al. (2007).
The $\sim$0.8 dex difference between the bulge and Sgr dSph [Al/Fe] ratios, at the same
[Fe/H] values, is remarkable.  In the Tinsley (1979) time-delay scenario, if the bulge
[Al/Fe] reflects the pure SNII ratio, then the Fe in the Sgr dSph is $\sim$85\%
SNIa material; however, if the bulge contains any Fe from SNIa, then the Sgr dSph SNIa Fe
fraction must be even greater.  Thus, the Sgr dSph iron-peak elements seem to be dominated
by SNIa material, and this system would be useful to compare with predicted
SNIa iron-peak yields.  In Figure~2 the RGB stars of the Sgr dSph and the bulge have
similar [Fe/H], and temperature, so non-LTE effects are unlikely to explain the
Al and Na abundance differences.

\subsection{S-Process Enhancements}

Mild enhancements in heavy s-process elements in the LMC supergiants, were first noted 
by Russell \& Bessell (1989), McWilliam \& Williams (1991) and Hill et al. (1995).
For the dwarf galaxies the small numbers of stars studied by Shetrone et al. (2001) showed
large dispersion, with possible slight enhancements in the mean heavy s-process noted.
However, firm detection of s-process enhancements in nearby dwarf galaxies were first seen
in Sgr dSph RGB stars, by Smecker-Hane \& McWilliam (2002), showing a steady increase in 
[La/Fe] with increasing [Fe/H], up to [La/Fe]=$+$1 dex at [Fe/H]$\sim$0.
Remarkably, the Ba~II lines were too saturated in the Sgr dSph for reliable abundance
measurement.  Figure~3 shows plot of [La/Eu] versus [La/H] adapted from McWilliam \& 
Smecker-Hane (2005a), indicating the dominance of the s-process, with a locus that indicates
halo composition plus at least 95\% s-process above [Fe/H]$\sim$$-$0.7 dex. 
Notably, the [Eu/Fe] ratio enhancement, at roughly $+$0.3 dex, is due to the
s-process, despite the fact the solar Eu is $\sim$95\% r-process.  An r-process
assignment can only be identified from the neutron-capture element abundance
ratios, not the ratio to iron.

\begin{figure}[ht]
\centering
\includegraphics[angle=0,width=13cm]{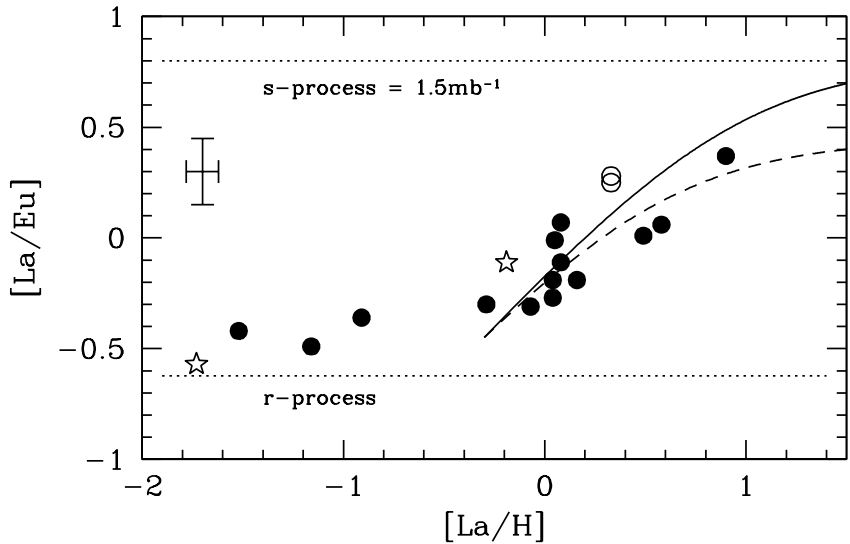}
\caption{[La/Eu]  versus [La/H]for the Sgr dSph, adapted from McWilliam \& Smecker-Hane (2005a).
Solid line shows locus of pure s-process added to an original composition; dashed line indicates
the locus of 95\% s-process plus 5\% r-process added to the original composition.}
\end{figure}

%The trend of [La/Eu] versus [La/H], where [La/H] serves as a 
%metallicity indicator.  At low metallicity the [La/Eu] ratio is typical of
%the halo, but as metallicity increases the ratio increases toward a value
%expected for strong s-processing, computed by Malaney (1987).  The loci
%in Figure~? show the path for an initial halo composition with the addition
%pure s-process, and a path for 95\% s-process plus 5\% r-process; the data
%demonstrates that no more than 5\% r-process contribution occurred during
%the late evolution of the heavy elements in Sgr dSph.

%An interesting point is that the [Eu/Fe] ratio in Sgr dSPh is enhanced by 
%$\sim$0.3 dex, even at solar metallicity; this is due to s-process Eu.  I
%note that while Eu is $\sim$95\% r-process in the sun, s-process enhancements
%in excess of 1 dex can increase the [Eu/Fe] ratio measurably.  Thus, it should
%not be assumed that large [Eu/Fe] ratios always indicates r-process enhancements;
%the neutron process must be identified by ratios among the heavy neutron-capture
%elements, not a ratio with iron.

The same authors found normal Y abundances (a light s-process element), giving a
high [La/Y], which measures the heavy/light, or [hs/ls], ratio; the enhanced
[hs/ls]$\sim$0.5 dex, occurs for Sgr dSph stars for [Fe/H] between $-$0.7 dex and the
solar value.

Busso et al. (1999) showed that high [hs/ls] ratios occur in metal-poor AGB s-processing.
At low metallicity the ratio of iron seed nuclei to neutrons is low, so each seed
nucleus captures many neutrons, thus pushing the synthesis to heavy nuclei.
From the [La/Y] ratio McWilliam \& Smecker-Hane (2005a)
concluded that the AGB s-processing took place in Sgr dSph stars with [Fe/H] near $-$0.6 or
below $-$1, which was much lower [Fe/H] than the solar metallicity Sgr dSph RGB stars studied.  Thus,
the Sgr dSph must have been polluted with ejecta from metal-poor AGB stars.

Similar s-process enhancements and patterns have been found in the LMC RGB stars by Pomp\'eia et al.
(2008), in the Fornax dSph by Letarte et al. (2010), and in the Ursa Minor dSph by 
Cohen \& Huang (2010); thus, this appears to be a general feature seen in the most metal-rich
stars of dwarf galaxies.

%An important point is that the light s-process elements are not noticeably enhanced,
%as demonstrated by the high [La/Y] ratios, and supported by the halo-like [Y/Fe]
%values.  This may seem confusing, until one recalls that the ratio of heavy to
%light s-process [hs/ls] is known to be metallicity-dependent (e.g. Busso et al. 2000??).
%At low metallicity there are many neutrons per iron-peak seed nuclei, so 
%each seed nucleus captures many more neutrons that in high metallicity environments,
%and pushes the pattern of enhancements to the heavier elements.
%McWilliam \& Smecker-Hane
%(2005) used this metallicity sensitivity of the [La/Y] ratio to conclude that the
%AGB stars responsible for the s-process in the Sgr dSph RGB stars had [Fe/H] near
%$-$0.6 or below $-$1.  The RGB stars with this high [La/Y] ratio were solar metallicity,
%so neither they, nor a companion, could have produced the measured s-process abundance
%pattern.  This indicates that the Sgr dSph RGB stars formed from material with enhanced
%[La/Y] ratios, suggesting that the entire galactic gas was polluted with ejecta from
%metal-poor AGB stars.

%These unusual neutron-capture abundance patterns have since been found in the LMC
%by Pompeia et al. (2008) and in the Fornax dSph by Letarte et al. (2010).  They
%appear to be a general phenomenon of stars in dwarf galaxies.

S-process enhancements have long been known for the
Galactic globular cluster, Omega~Cen (e.g., Vanture, Wallerstein \& Brown 1994; 
Smith et al. 2000).  I note that a plot of [La/Eu] versus [La/H] from the data of
Johnson \& Pilachowski. (2010) shows, for nearly all stars, a locus that is consistent 
with the addition of pure s-process material to the composition of the oldest stars in 
Omega~Cen, similar to Sgr dSph (Figure~3 here) from McWilliam \& Smecker-Hane (2005a).

Together with similar Cu and Mn abundances, this chemical similarity with the Sgr dSph 
%(e.g., Cunha et al. 2???; Cunha et al. 2010; McWilliam et al. 2003)
suggests a common history, and that Omega~Cen may be the core of an accreted dwarf galaxy.  
The main chemical difference is that Omega~Cen possesses halo-like, enhanced, [$\alpha$/Fe] ratios.
I find this difference both puzzling and fascinating, because it is apparently inconsistent with 
the Tinsley time-delay scenario; thus, the $\alpha$-enhancements in Omega~Cen may provide a critical
clue for understanding chemical evolution.

The s-process abundance patterns in these dwarf systems indicates significant
nucleosynthesis by metal-poor AGB stars, which McWilliam \& Smecker-Hane (2005a) suggested
resulted from leaky box chemical evolution, in which the MDF possessed a larger
fraction of metal-poor stars than the solar neighborhood.  Initially, these galaxies
would have formed early, metal-poor, populations proportional to the mass of gas,
but by late times a large fraction of the gas had leaked-out, such that the ejecta
from the now relatively large population of old, metal-poor, AGB stars dominated neutron-capture
element abundance pattern of the late-time gas.   Presumably, leaky-box chemical evolution must be
quite general and apply to many dwarf galaxies, and it seems reasonable that the
amplitude of the effect may be greater for lower-mass galaxies.

\subsection{Manganese and Copper}

The solar neighborhood (i.e., halo, thin and thick disk populations) trend for
[Cu/Fe] shows a sub-solar plateau, near [Cu/Fe]=$-$0.6 dex, below [Fe/H]$\sim$$-$1.5;
above that the [Cu/Fe] ratio increases approximately linearly with [Fe/H], toward
the solar composition at [Fe/H]=$-$0.5, and remains constant thereafter
(see Mishenina et al. 2002; Simmerer et al. 2003).  It appears that the thin disk
[Cu/Fe] ratio is constant, while there is an approximately linear [Cu/Fe] trend with 
[Fe/H] for the transition from halo to thick disk populations.

McWilliam \& Smecker-Hane (2005b) found severe [Cu/Fe] deficiencies, up to nearly 0.6 dex,
in the Sgr dSph for [Fe/H] above $-$0.8 dex.  In the same month Geisler et al. (2005) found
similar [Cu/Fe] deficiencies in the Sculptor dwarf galaxy, but at lower [Fe/H] than in the
Sgr dSph.  These deficiencies were reminiscent of those found in Omega~Cen by Cunha et al. (2002).
More recently, Pompeia et al. (2008), Figure 3 here, found large [Cu/Fe] deficiencies in LMC stars
with [Fe/H] above $-$1, almost as if the low [Cu/Fe] ratios in the halo has been extended up to the
highest [Fe/H] in the LMC.  Carretta et al. (2010) have confirmed the low [Cu/Fe] ratios in the 
Sgr dSph.  Included without discussion in their paper, the [Cu/Fe] ratios of Cohen \& Huang (2010) 
for the UMi dSph are deficient by $\sim$0.4 dex, relative to the Milky Way.  Clearly,
Cu deficiencies are a common, perhaps ubiquitous, signature of dwarf galaxies.

\begin{figure}[ht]
\centering
\includegraphics[angle=0,width=10cm]{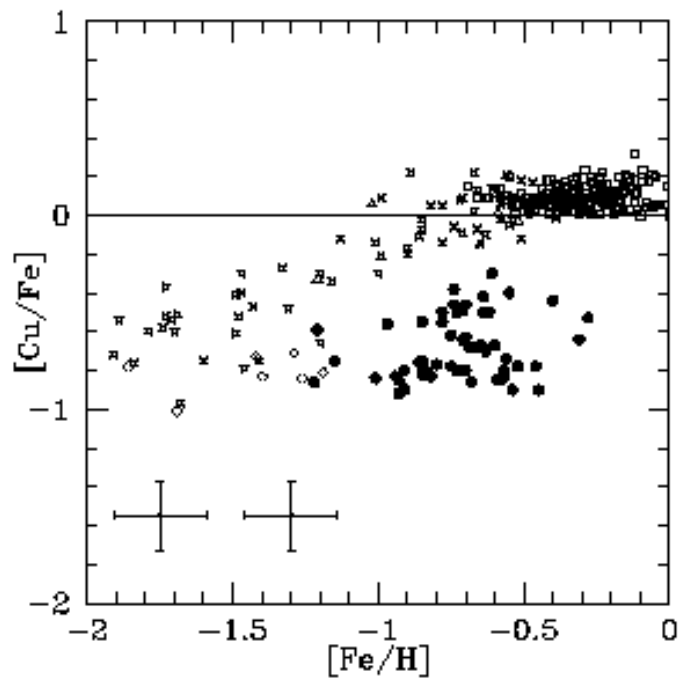}
\caption{[Cu/Fe] for the LMC (filled circles) taken from Pomp\'eia et al. (2008); similar trends are
also seen in Sgr, Sculptor, and UMi dSphs.  Other symbols represent Milky Way stars.}
\end{figure}

Bisterzo et al. (2004) concluded that the {\em weak sr-process} in massive stars is the
principle source of Cu in the Galaxy, occurring during convective
shell C burning of SNII progenitors, and that SNIa and AGB stars do not produce significant
quantities of Cu.  The metallicity-dependence
of the weak sr-process explains the positive slope of [Cu/Fe] with [Fe/H], while
the production of iron from SNIa explains the flattening-out of the slope in the thin
disk.

The Cu deficiencies in dwarf galaxies indicates a paucity of high-mass stars in
these systems, although the metallicity-dependence could play a role.
This lack of material from massive stars is consistent with the prolonged star formation and
leaky box scenario suggested by the abundances of the $\alpha$ and s-process elements.  By 
this same argument, one would
expect that bulges and giant ellipticals, or systems with high SFR and/or an excess of 
high-mass stars, should possess [Cu/Fe] enhancements, relative to the solar neighborhood.

In the solar neighborhood  [Mn/Fe] 
shows approximately constant deficiency, near $-$0.4 dex, below [Fe/H]=$-$1, but
increases roughly linearly with increasing [Fe/H] to [Mn/Fe]=0.0 at solar metallicity
(Sobeck et al. 2006; Gratton 1989).  Because this trend looks like the mirror image of the
[$\alpha$/Fe] trend
with [Fe/H], Gratton (1989) suggested that Mn is over-produced in SNIa.  In this scenario
the time-delay for the onset of SNIa would be responsible for the increase in [Mn/Fe] above
[Fe/H]$\sim$$-$1.  If Gratton's suggestion is correct, the time delay scenario of Tinsley
(1979) and MB90 would predict that [Mn/Fe] is enhanced in 
low SFR dwarf galaxies, but deficient in high SFR systems, such as bulges and elliptical
galaxies.

Measurements of [Mn/Fe] abundance ratios in Sgr dSph RGB stars by McWilliam et al. (2003)
showed a deficiency relative to the solar neighborhood
trend by $\sim$0.2 dex, in stark contrast to the expectations if Mn is over-produced by
SNIa.  Similar Mn deficiencies have been seen in Omega~Cen by Cunha et al. 
(2010), but Carretta et al. (2010) did not confirm this finding in their sample 
of Sgr dSph stars.  The low [Mn/Fe] ratios for LMC supergiants, by Hill et al. (1995), are
similar to the deficiencies seen in the Sgr dSph; more work on LMC [Mn/Fe] ratios is warranted.

McWilliam et al. (2003) also found that the [Mn/Fe] trend with [Fe/H]
in the Galactic bulge is similar to the solar neighborhood; assuming that the bulge formed
on a short timescale, this is also inconsistent with the expectations if SNIa over-produce
Mn, since a Mn deficiency would be predicted for the bulge.  
Arnett (1971) concluded that the Mn yield is metallicity-dependent in SNII; McWilliam et al. (2003)
speculated that this metal-dependence should also apply to SNIa, which also synthesize the
iron-peak elements.  Thus, low [Mn/Fe] ratios in Sgr dSph could be explained by metal-poor SNIa,
which might be expected to accompany the metal-poor AGB population.

%--------------------------------------------
%Other chemical peculiarities seen in the Sgr dSph, found by McWilliam \& Smecker-Hane
%(2005) and McWilliam et al. (2003?), include deficiencies of Cu and Mn below the
%sub-solar [Mn/Fe] and [Cu/Fe] trends in the solar neighborhood.  The low [Mn/Fe]
%ratios in Sgr dSph ran contrary to a previous proposal for Mn production, suggested
%by Gratton (19??).  In this scenario Gratton suggested that the deficiency in
%
%Cu deficient in Sgr dSPh and Omega Cen and LMC
%
%MN in SGR and OMEGA CEN
%------------------------

\section{A Qualitative Model for Dwarf Galaxy Evolution}

The chemical properties of the dwarf galaxies, outlined above, can be 
qualitatively explained in a model of prolonged chemical enrichment with
on-going gas-loss, or leaky-box chemical evolution.

The low [$\alpha$/Fe] trends, the Na and Al deficiencies, and the low [Cu/Fe]
ratios at higher metallicities, all indicate low
SNII/SNIa ratios in dwarf systems, consistent with low SFRs and prolonged evolution.  
In particular, the comparison of [Al/Fe] from Sgr dSph to thin disk to bulge
suggests that the Sgr dSph iron-peak material near solar [Fe/H] is $\sim$85\% 
SNIa.  Thus, the dwarf galaxies are good places to look for the signature of SNIa
nucleosynthesis.

These prolonged formation timescales
are supported by the significant enhancements of s-process material, produced
by relatively low mass AGB stars with long main sequence lifetimes.  In the case
of the Sgr dSph the ages and metallicities of its associated
globular clusters also indicate a long formation timescale.

The low mean metallicities and high [hs/ls] s-process ratios provide strong evidence 
for significant mass-loss from the dwarf galaxies during their evolution.  Without such
a leaky box chemical evolution the
high [hs/ls] ratios from the metal-poor AGB stars would have been overwhelmed by
low [hs/ls] ratios at higher [Fe/H], and the mean metallicities would be higher.

In a leaky box at early times a significant metal-poor population is formed, but
much of the gas is lost from the galaxy during subsequent evolution, until at late times
there is very little gas to form new generations of stars.  Thus, there are very
few SNII at late times to produce higher $\alpha$/Fe, Al/Fe, Na/Fe, Cu/Fe ratios, and
few high metallicity AGB stars producing low [hs/ls] s-process ratios.  However, by late
times the old, metal-poor, population
AGB stars eject significant amounts of gas and s-process elements, and thus these dominate the
abundance pattern of the neutron-capture elements of the younger, more metal-rich, population.

In this scenario an important question is: where do the iron-peak elements come from
at late times?  The answer is that they likely come from low-mass SNIa systems,
which have prolonged main-sequence lifetimes, and are relatively metal-poor, and should accompany
the old, metal-poor, AGB population.  The low
metallicity of the SNIa explains the measured low [Mn/Fe] ratios in the more metal-rich
stars of the Sgr dSph and Omega~Cen, since Mn yields are expected to be metallicity-dependent.

Detailed chemical evolution models are required to constrain the amount of gas lost
from these dwarf galaxies.  
Other phenomena to be modelled include the retention of ejecta from various sources, such as
AGB stars, SNIa, and SNII, which have very different kinetic energies; also, the effects of dark
matter and cooling (i.e., metallicity) on gas retention.
The models would be very helpful for understanding the evolution of these galaxies and to constrain
the sites of nucleosynthesis for various elements, and constraining stellar yields.  It would be
helpful to have a set of predicted abundance ratios for dwarf galaxies from such detailed models,
as a function of some mass-loss parameter.  So far, most of the abundance anomalies found in dwarf
galaxies were observed, not predicted.

%Other issues to be modelled include selective retention of low-velocity ejecta from AGB
%stars, and the effect of cooling on
%
%Another
%possibility is that low-mass galaxies selectively retain more material from AGB stars than 
%SNIa or SNII, due to the velocity of the ejecta.
%
%the role of selective retention of ejecta by low mass galaxies.  It may be that the strong
%s-process enhancements and high [hs/ls] pattern results from the retention of AGB material
%by the shallow gravitational potential of these galaxies, where SNIa and SNII material
%%are preferentially ejected, due to their large kinetic energies.  The temperatures of the
%ejecta and their ability to cool (i.e., metallicity) may also affect retention and
%abundance patterns in dwarf galaxies.

Of significant interest is the role of the IMF in dwarf galaxy chemical evolution.
In particular, a major question for understanding star formation in galaxies is whether
the IMF slope is everywhere the same, but with reduced high-mass end cutoffs in dwarf
systems.  It is possible to probe the IMF with various element ratios that are sensitive to 
stellar mass (e.g., [Mg/Ca], [C/O]), although the interpretation may
require modelling.  Comprehensive CNO abundance results for dwarf galaxies, and
other element ratios, such as [Rb/Zr] and D/H in these systems would provide interesting
tests of the paradigm outlined here.

%\section{...}

\end{document}